\documentclass[a4paper,11pt]{article}
\pdfoutput=1 

\usepackage{jheppub} 

\usepackage[T1]{fontenc} 
\usepackage{slashed}
\usepackage{graphicx}
\usepackage{subcaption}
\usepackage{color,graphicx,epsfig}
\usepackage{amsmath} 

\DeclareUnicodeCharacter{2212}{-} 

\title{\boldmath Hierarchical High-Point Energy Flow Network for Jet Tagging}
\author[a,b]{Wei Shen,}
\author[c]{Daohan Wang,}
\author[a,b]{Jin Min Yang}

\affiliation[a]{CAS Key Laboratory of Theoretical Physics, Institute of Theoretical Physics, Chinese Academy of Sciences, Beijing 100190, P. R. China}
\affiliation[b]{School of Physics Sciences, University of Chinese Academy of Sciences,  Beijing 100049,  P. R. China}
\affiliation[c]{Department of Physics, Konkuk University, Seoul 05029, Republic of Korea}

\emailAdd{shenwei@itp.ac.cn}
\emailAdd{wdh9508@gmail.com}
\emailAdd{jmyang@itp.ac.cn}

\abstract{Jet substructure observable basis is a systematic and powerful tool for analyzing the internal energy distribution of constituent particles within a jet. In this work, we propose a novel method to insert neural networks into jet substructure basis as a simple yet efficient interpretable IRC-safe deep learning framework to discover discriminative jet observables. The Energy Flow Polynomial (EFP) could be computed with a certain summation order, resulting in a reorganized form which exhibits hierarchical IRC-safety. Thus inserting non-linear functions after the separate summation could significantly extend the scope of IRC-safe jet substructure observables, where neural networks can come into play as an important role. Based on the structure of the simplest class of EFPs which corresponds to path graphs, we propose the Hierarchical Energy Flow Networks and the Local Hierarchical Energy Flow Networks. These two architectures exhibit remarkable discrimination performance on the top tagging dataset and quark-gluon dataset compared to other benchmark algorithms even only utilizing the kinematic information of constituent particles.}


\begin{document} 
\maketitle
\flushbottom
\newpage

\section{Introduction}
Identifying deviations from the Standard Model predictions in a vast amount of high-energy event data generated at a collider like the LHC is an important task in the search of new physics.  One of the most crucial objects in analyzing high-energy collision products is jets, which are collimated sprays of outgoing particles produced in a high energy collision. Jet substructure refers to a series of well-defined physical observables that measure the particle and energy distribution within a jet, reflecting the internal radiation pattern. Identifying novel jet patterns helps to separate signals of interest and provides a quantitative understanding of the underlying mechanisms.

Jet substructure basis is a set of systematic and (over-)complete basis for jet observables. Given that a jet typically consists of hundreds of constituent particles, there is a huge number of possible observable combinations that can serve as a complete basis for such a multi-body system. In practice, there are some restrictions on selecting robust and well-defined observables. To guarantee a good definition in perturbative QCD calculation as well as the robustness to experimental resolution eﬀects, any jet substructure observable should be both infrared and collinear safe (IRC-safe) and Lorentz invariant. Previous studies have proposed a generalized form known as calorimeteric-correlators (C-correlators) based on energy-weighted summation for direction-related features of all particles~\cite{whatisjet}, which maintains both IRC-safety and permutation symmetry. 
Furthermore, any Lorentz invariant jet observable can be decomposed into a combination of Lorentz invariants of particle pairs. For highly-boosted and narrow jets, the requirement of Lorentz invariance can be relaxed to $\text{SO}(2)$-rotational symmetry around the jet axis~\cite{gurari2011classification}, and then the angular distance $R_{ij}$ between particles $i$ and $j$ are often used to approximate pair-wise Lorentz invariant features. Under these conditions, most of the energy-flow based substructure basis variables like N-subjettiness~\cite{1011.2268}\cite{1108.2701}, energy correlation functions (ECFs)~\cite{1305.0007}, generalized energy correlation functions (ECFGs)~\cite{1609.07483} and energy-flow polynomials (EFPs)~\cite{efps} have been proposed. They focus on magnifying significant difference in radiation patterns inside a jet, especially the multi-prong structure of a highly-boosted jet~\cite{Marzani_2019}. Besides these substructure basis, there are also several physically defined observables, such as the jet mass, angularities~\cite{Berger_2004} and planar flow~\cite{Almeida_2009}. Due to the large amount of data and the great freedom of features selection, machine learning is playing an increasingly important role in jet analysis in high energy collisions (for reviews on machine learning in high energy physics, see, e.g., ~\cite{Albertsson:2018maf,Abdughani:2019wuv,Plehn:2022ftl,Cheng:2022idp}).

Traditional machine learning methods, such as the boost decision trees (BDTs)~\cite{Roe_2005}, are widely employed to analyze high-level jet features like jet mass and N-subjettiness, and linear regression is also used to find discriminative jet observables by determining the coefficients of jet substructure bases like EFPs~\cite{efps}. 
Over the last decade, there has been a widespread adoption of deep learning techniques to improve the performance of jet tagging~\cite{Larkoski:2017jix,Larkoski_2020,thais2022graph}. Recently, there is an increasing emphasis on developing network architectures that prioritize both infrared and collinear (IRC) safety and physical interpretability for collision objects, rather than solely optimizing for downstream task performance. 
More recently, the point cloud jet representation, which treats the constituent particles within a jet as points in a point cloud, has gained significant attention and several deep learning architectures are proposed based on the point cloud representation, including Energy Flow Network~\cite{efn}, Energy-weighted Messaging Passing (EWMP) Neural Network~\cite{ewmp}, ParticleNet~\cite{particlenet}, ABCNet~\cite{Mikuni:2020wpr} and LorentzNet~\cite{Gong:2022lye}. The Energy Flow Network is an energy-weighted deep set network serving as an IRC-safe backbone model on the point cloud representation of jets~\cite{pointnet}. It parameterizes angular filters with trainable neural networks and has been successfully tested on many jet tagging tasks~\cite{efn}. Meanwhile, the Energy-weighted Messaging Passing (EWMP) Neural Network ~\cite{ewmp} maps kinematic features of particles to nodes and distances between particles to edges in a graph, which highlights that defining only the Radius Neighbor can make the algorithm IRC-safe, and compares the tagging performance at different aggregation radius settings. Additionally, the ParticleNet~\cite{particlenet} architecture has been proposed to process the local structure of jets permutations invariantly by aggregating the information of K-Nearest Neighbor (KNN) through a convolution block while the ABCNet employs the attention mechanism to extract the local structure of jets. The LorentzNet puts greater emphasis on integrating inductive biases derived from physics principles into its architectural design, employing a highly efficient Minkowski dot product attention mechanism. All these architectures exhibit excellent performance when applied to top tagging and quark/gluon discrimination benchmarks. Nevertheless, it is worth noting that only the EFN and the EWMP architectures maintain the IRC-safety.

In this work, we focus on inserting neural networks into jet substructure bases in an IRC-safe and rotation-invariant way to discover interpretability and discriminative jet observables from a vast amount of simulation data. 
The article is organized as follows. In Section 2, we give a brief introduction for some representative jet substructure observable bases, and propose to use different order of Legendre polynomials as functions of angular distance between particle pairs to achieve numerical stability. As an example, we present the distributions of 4-point Legendre-based path graph energy-flow polynomials for both top jets and QCD jets, spanning across different parameter settings. We also show how to determine the specific form of the 2-point energy-flow polynomials by employing linear regression on the coefficients of Legendre polynomials as an illustrative example. Most importantly, we point out that rewriting EFPs in a hierarchical manner would greatly reduce the computation complexity and allow for the insertion of neural network. In Section 3, we introduce the Hierarchical Energy Flow Networks and Local Hierarchical Energy Flow Networks which follow the path-graph structure, and provide comprehensive details about the model implementation. In Section 4, we present numerical results obtained for top tagging task and quark/gluon discrimination task, respectively. Finally, our conclusion are presented in Section 5.

\section{Hierarchical Energy Flow for Observables}
\subsection{Jet Substructure Observable Basis}
\begin{figure}
    \centering
    \includegraphics[width=16cm]{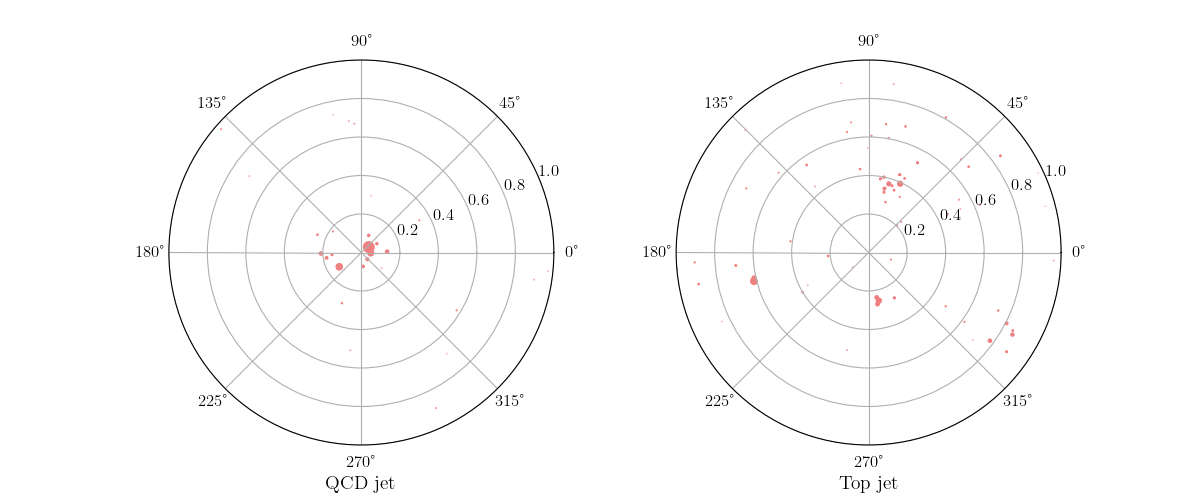}
    \caption{The visualizations of point cloud representation for a QCD jet and a top jet in polar coordinate. The jet radius is set to 1 and each red point denotes the position of a constituent particle within the jet.}
    \label{fig:top vs qcd}
\end{figure}

To guarantee a good definition in perturbative QCD calculation, any jet substructure observable should be both infrared and collinear safe (IRC-safe) and Lorentz invariant. These properties can be achieved by expressing an observable as a linear combination of calorimetric correlators (C-correlators)~\cite{whatisjet} with a general form
\begin{equation}
    C_{N}^{f_{N}}=\sum_{i_{1}=1}^{M}...\sum_{i_{N}=1}^{M}E_{i_{1}}...E_{i_{N}}f_{N}(\{\hat{p}_{i_{1}}, ..., \hat{p}_{i_{N}}\}),
    \label{C-correlators}
\end{equation}
where $E_{i}$ and $\hat{p}_{i}$ are respectively energy (or energy fraction) and direction of the $i$-th constituent particle within the jet, $M$ is the total number of the constituent particles inside the jet, $N$ is any positive integer, and $f_N$ is any suﬃciently smooth permutation invariant function of its $N$ arguments. The multi-particle energy correlator forms are naturally derived from quantum field theory. The function $f_{N}$ responsible for maintaining the calorimetric continuity depends solely on the directions of particles. Numerous efforts have been made to define various forms of function $f_{N}$ as variants of the energy-flow based observables. The Deep Set Theorem~\cite{zaheer2018deep} indicates that the functions operating on point sets should be permutation invariant to the order of objects in the set. By further enforcing the IRC-safety, any observable $\mathcal{O}$ of a variable-length point set can be approximated arbitrarily well by~\cite{efn}
\begin{equation}
    \mathcal{O}(\{p_{1},...,p_{M}\})=F(\sum_{i=1}^{M}z_{i}\Phi(\hat{p}_{i})).
    \label{efn_approx}
\end{equation}
Fig.\ref{fig:top vs qcd} presents a visualization of point cloud representation of a jet in polar coordinate. To enhance the flexibility of the representation, the Energy Flow Networks~\cite{efn} have been introduced, which directly replace the fixed form of expansion polynomials with parameterized and learnable filters $\Phi$ in neural networks. The universal approximation theorem of neural networks supports this formulation, making it more adaptable and versatile. There are also some attributes that utilize a systematic and complete basis $\Phi^{l}$ for angular directions to decompose complete features of the jet global observables, including Zernike polynomials~\cite{gurari2011classification} and spherical harmonics ~\cite{PhysRevLett.41.1581}.
 
For a jet with N-prong substructure, it is also efficient to define the N-subjettiness~\cite{1011.2268} to measure the energy radiation alignment with $N$ candidate subjets: 
\begin{equation}
    \tau_{N}^{\beta}= \sum_{i}E_{i}\min\{\Delta R_{i, 1}^{\beta}, \Delta R_{i, 2}^{\beta}, ..., \Delta R_{i, N}^{\beta}\},
\end{equation}
where $\Delta R_{i, j}$ represents the angular separation between $i$-th and $j$-th candidate subjets in the angular plane. The parameter $\beta$ acts as the angular exponent, enabling the capture of the $3M-4$ dimensional representation of the $M$-body phase space~\cite{mbody}. 
To determine the axes of the fixed number of candidate subjets, the exclusive-$k_{T}$ algorithm, similar to the weighted $k$-mean algorithm, is used to implement the minimization process. 
Additionally, several specialized approaches for evaluating $N$-prong structures of a jet are proposed. One method is the $N$-point energy correlation functions (ECFs) ~\cite{1305.0007}, which are defined as 
\begin{equation}
    \text{ECF}(N,\beta) = \sum_{i_{1}<i_{2}<...<i_{N}\in J}(\prod_{a=1}^{N}E_{i_{a}})(\prod_{b=1}^{N-1}\prod_{c=b+1}^{N}\theta_{i_{b}i_{c}})^{\beta},
\end{equation}
where $\theta_{ij}$ is the angular distance between particles $i$ and $j$. Another method, known as the generalized energy correlation functions (ECFGs) ~\cite{1609.07483} use minimizing steps in $N$-subjettiness and compare its superiority. Moreover, the energy flow polynomials (EFPs)~\cite{efps} provide a systematic and organized method for calculating a more (over)complete basis of jet substructure, with a topological correspondence to multi-graphs:
\begin{equation}
    \text{EFP}_{G} = \sum_{i_{1}=1}^{M}...\sum_{i_{N}=1}^{M} E_{i_{1}}...E_{i_{N}}\prod_{(k,l)\in G}\theta_{i_{k}i_{l}}.
\end{equation}
These EFPs have been demonstrated to compose existing physics-inspired observables like jet mass, planar flow~\cite{Almeida_2009} and the above-mentioned ECFs (corresponding with complete graph). Compared with the moment expansion representation including jet image mentioned above, those higher-point energy flow bases are more complete and show clearer correspondence to the physically well-defined observables. Note that the use of the Euclidean distance metric in the energy flow basis indicates the rotation symmetry of a single jet. However, when we generalize this to the construction of global event features, like jet pull ~\cite{Gallicchio_2010} and other color flow variables, with higher-point energy flow bases, directional functions $f_{N}$ of rapidity and azimuth directions should be defined separately. 

The systematic comparison studies demonstrate the discrimination performance of the energy correlation functions for $N$-prong jets under various angular exponent $\beta$~\cite{1305.0007}. In this study, we propose a method to modify $\theta_{ij}^\beta$ in the EFPs to a complete and orthogonal basis, such as Legendre polynomials, to parameterize and vectorize the angular distance function relationship. This approach allows us to effectively combine the influence of different angular exponents. Furthermore, we would like to point out that it may not be necessary to strictly adhere to the formula where the function $f_{N}$ in the C-correlator is solely dependent on particle directions to ensure IRC safety.
By liberating from this constraint, the possible formulations of jet observables are greatly expanded to beyond the linear combination of C-correlators.

\subsection{Hierarchical Energy Flow Functions}
Now we consider a naive 2-point energy correlation function $\text{ECF}(2, \beta)$ 
\begin{equation}
\text{ECF}(2,\beta)=\sum_{i,j}^{M}z_{i}z_{j}R_{ij}^{\beta}.
\end{equation}
We could generalize the function of angular distance between two particles $R_{ij}$ as any function $f(R_{ij})$, and then expand it with orthogonal polynomials like Legendre polynomials $\text{P}_{\beta}(\theta_{ij})$. We set $\theta_{ij}=R_{ij}/R_{0}-1$ to make $\theta_{ij}$ in the domain $[-1,+1]$ to keep numerical stability. Then the Legendre-based 2-point energy correlation function $\text{ECF}(2,\text{P}_{\beta})$ can be written as 
\begin{equation}
\text{ECF}(2,\text{P}_{\beta}) = \sum_{i,j}^{M}z_{i}z_{j}\text{P}_{\beta}(\theta_{ij}).
\end{equation}
However, it has been pointed out that the higher-point energy correlation functions pose challenges due to their power-growth computational complexity~\cite{1305.0007}. The computational complexity of $N$-point Energy Correlation Functions (ECFs) grows significantly, denoted as $\mathcal{O}(M^{N})$, where $M$ represents the number of particles inside the jet. Fortunately, some special $N$-point Energy Flow Polynomials (EFPs) associated with tree graphs could be computed by utilizing the Variable Elimination (VE) algorithm~\cite{efps} with a reduced complexity of $\mathcal{O}(M^{2})$. Here we select a subset from those special EFPs corresponding to the path graphs, which are the simplest example of tree graphs. Therefore, the path-graph corresponding EFPs can be expressed in a hierarchical summation form. For example, the 2-pint, 3-point and 4-point path-graph Hierarchical Energy Flow Polynomials (HEFPs) are defined as
\begin{eqnarray}
&&    \text{HEFP}(2,\beta)=\sum_{i}^{M}z_{i}(\sum_{j}^{M}z_{j}\text{P}_{\beta}(\theta_{ij})),\\
&&    \text{HEFP}(3,\{\beta_{1},\beta_{2}\})=\sum_{i}^{M}z_{i}(\sum_{j}^{M}z_{j}\text{P}_{\beta_{2}}(\theta_{ij})(\sum_{k}^{M}z_{k}\text{P}_{\beta_{1}}(\theta_{jk}))),\\
&&    \text{HEFP}(4,\{\beta_{1},\beta_{2},\beta_{3}\})=\sum_{i}^{M}z_{i}(\sum_{j}^{M}z_{j}\text{P}_{\beta_{3}}(\theta_{ij})(\sum_{k}^{M}z_{k}\text{P}_{\beta_{2}}(\theta_{jk})(\sum_{l}^{M}z_{l}\text{P}_{\beta_{1}}(\theta_{kl}))))
\end{eqnarray}
The compact formulas of path-graph $(N+1)$-point HEFPs are
\begin{equation}
    \text{HEFP}(N+1,\{\beta_{1},...,\beta_{N}\})=\sum^{M}_{i_{1},i_{2},...,i_{N+1}}(\prod_{a=1}^{N+1}z_{i_{a}})(\prod_{b=1}^{N}\text{P}_{\beta_{i_{b}}}(\theta_{i_{b}i_{b+1}})).
\end{equation}
They could be seen as an energy-weighted summation of $N$-point particle features:
\begin{equation}
    \text{HEFP}(N+1, \{\beta_{1},...\beta_{N}\}) = \sum_{i}^{M}z_{i}\hat{p}^{N}_{i}(\{\beta_{1},...\beta_{N}\}),
\end{equation}
where $\hat{p}_{i}^{N}(\{\beta_{1},...\beta_{N}\})$ are obtained through recursive relations as
\begin{equation}
    \hat{p}^{t+1}_{i}(\{\beta_{1},...\beta_{t+1}\}) = \sum_{j}^{M}z_{j}\hat{p}_{j}^{t}(\{\beta_{1},...\beta_{t}\})\text{P}_{\beta_{t+1}}(\theta_{ij}),
\end{equation}
with $(t+1)$-point particle features $\hat{p}^{t+1}_{i}$ being an energy-weighted summation of all $t$-point particle features $\hat{p}^{t}_{j}$ multiplied by a pair-wise distance related factor $P_{\beta_{t+1}}(\theta_{ij})$. 

The above path-graph HEFPs are linear combinations of multi-graph corresponding EFPs. Nevertheless, the integration of Legendre polynomials in our approach offers several crucial advantages. These polynomials serve as a powerful and flexible tool for representing any complex angular distance functions, $f(\theta_{ij})$, in a compact and numerically stable manner. By expanding $f(\theta_{ij})$ using Legendre polynomials $\text{P}_{\beta}(\theta_{ij})$, we can effectively capture the essential features of the angular distribution while controlling the model complexity. Notably, the orthogonality property of Legendre polynomials guarantees that each coefficient in the expansion corresponds to a distinct feature, enhancing the interpretability of the model. Additionally, by transforming $R_{ij}$ to the range of $\theta_{ij}\in[-1, +1]$, we effectively restrict the absolute value of Legendre polynomials $\text{P}_{\beta}(\theta_{ij})\in [0, 1]$, further improving the model's robustness.

\begin{figure*}[ht]
\begin{center}
\includegraphics[width=7.5cm]{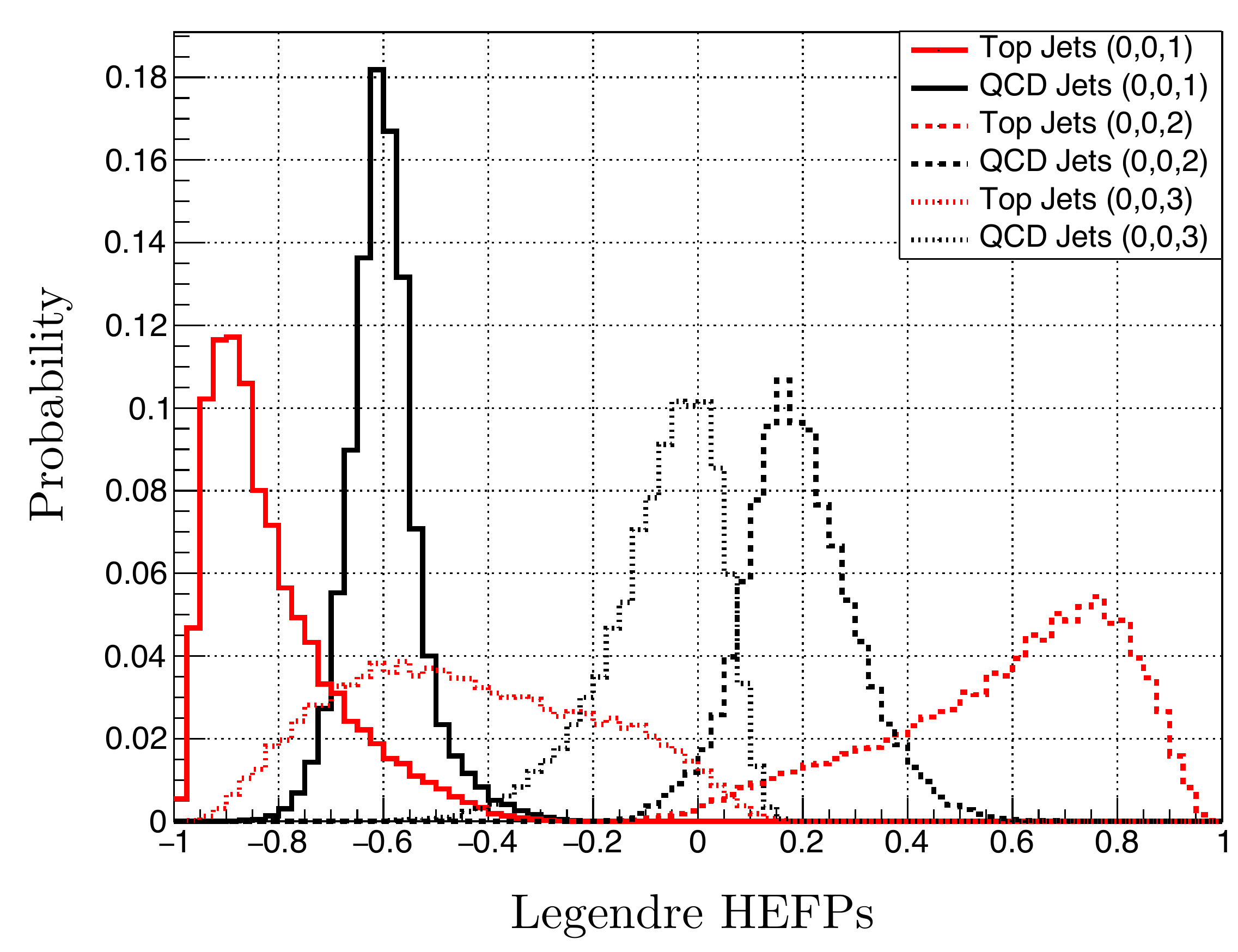}
\includegraphics[width=7.5cm]{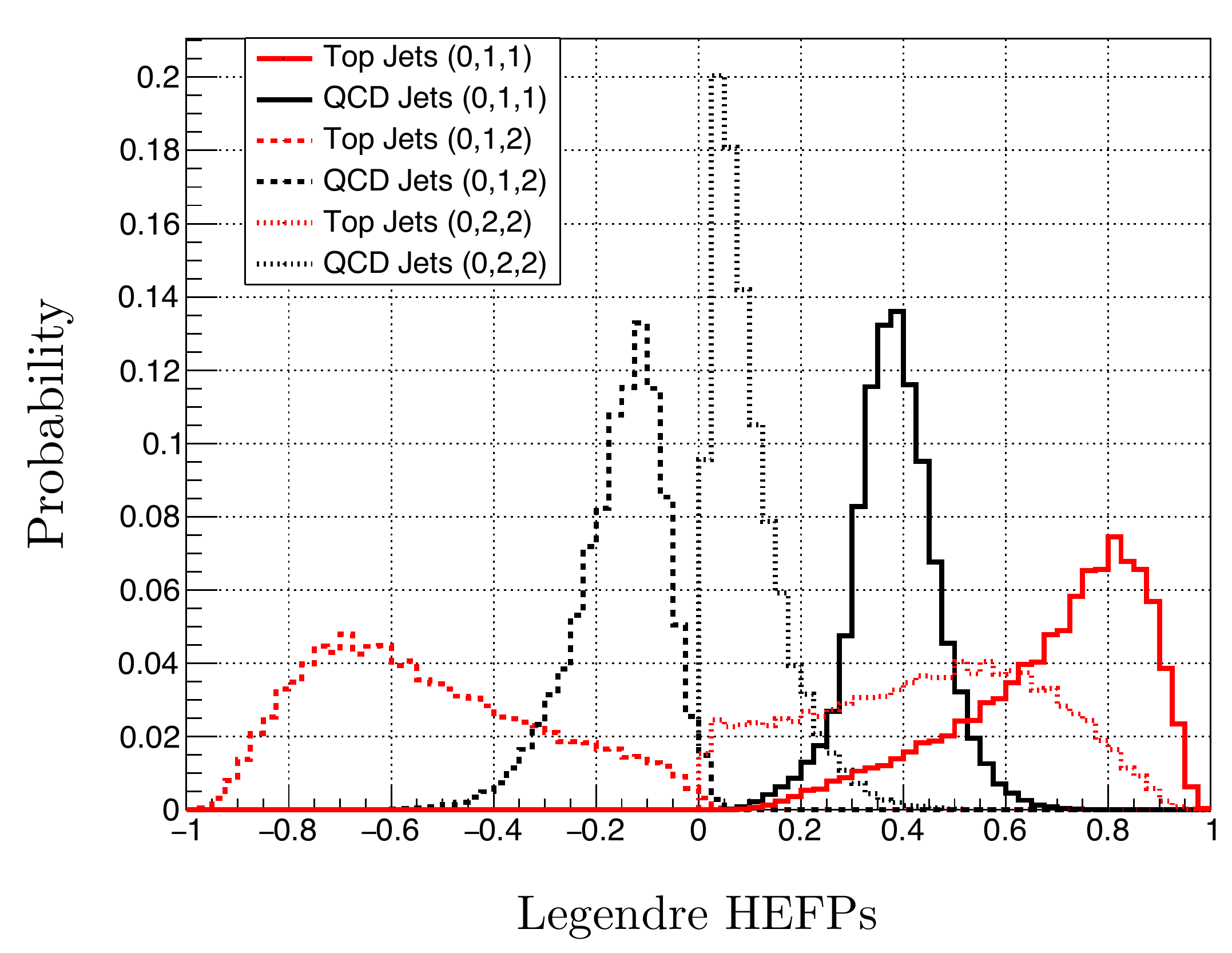}
\includegraphics[width=7.5cm]{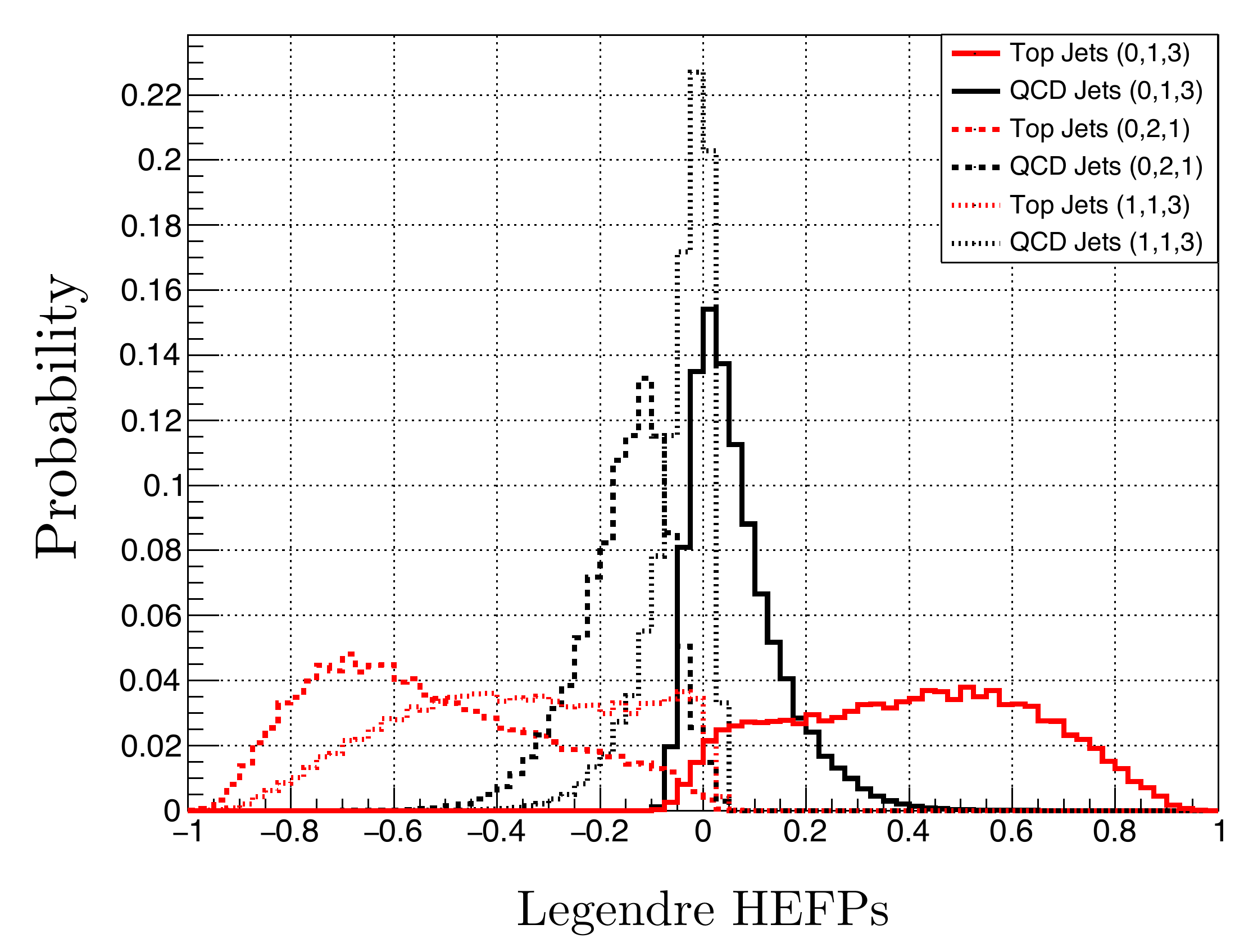}
\includegraphics[width=7.5cm]{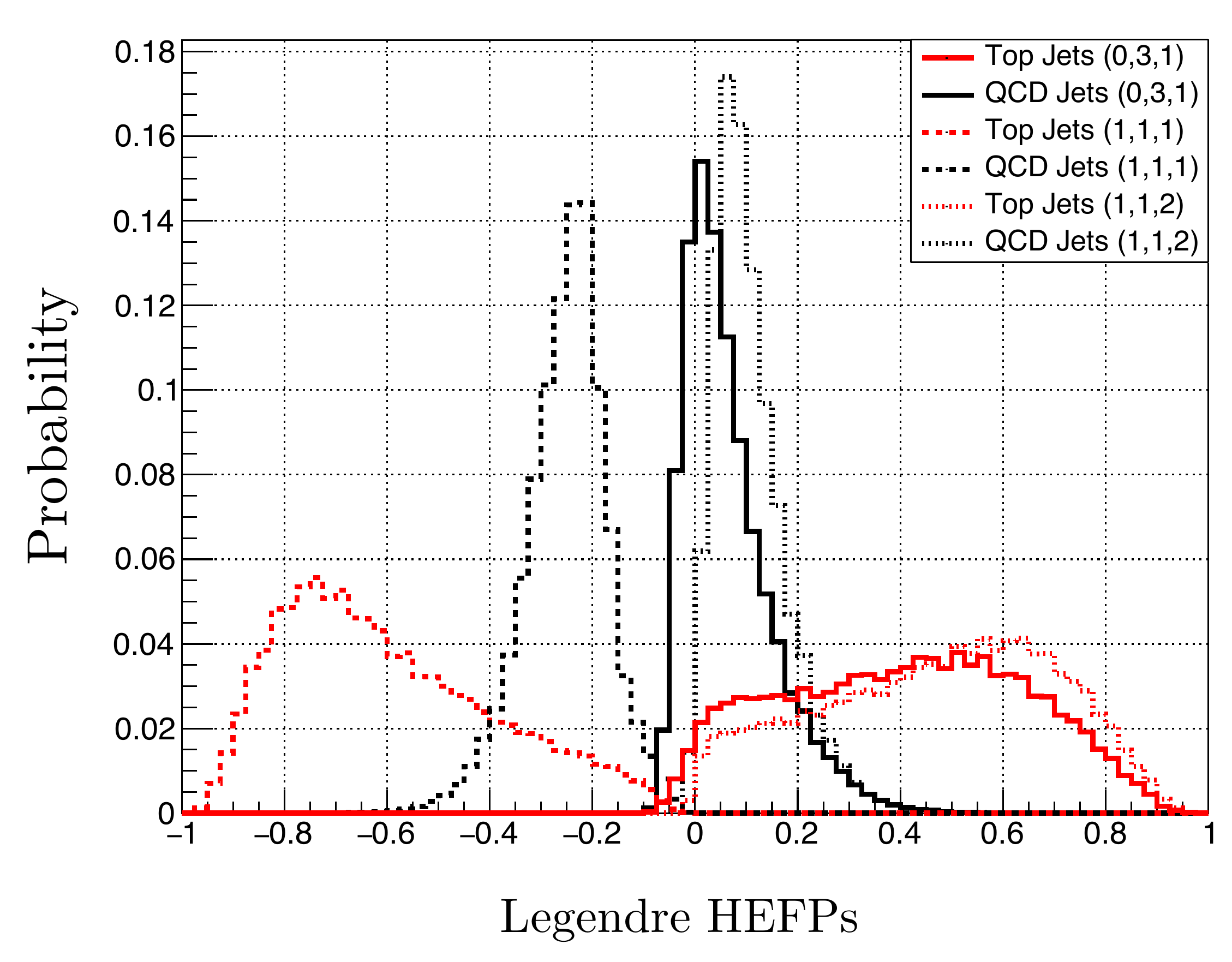}
\caption{Distributions of 4-point Legendre Hierarchical Energy Flow polynomials \text{HEFP}(4, $\{\beta_{1},\beta_{2},\beta_{3}\}$) under various settings of $(\{\beta_{1},\beta_{2},\beta_{3}\})$ for top jets and QCD jets.}
\label{distributions1}
\end{center}
\end{figure*}

\begin{figure*}[ht]
\begin{center}
\includegraphics[width=7.5cm]{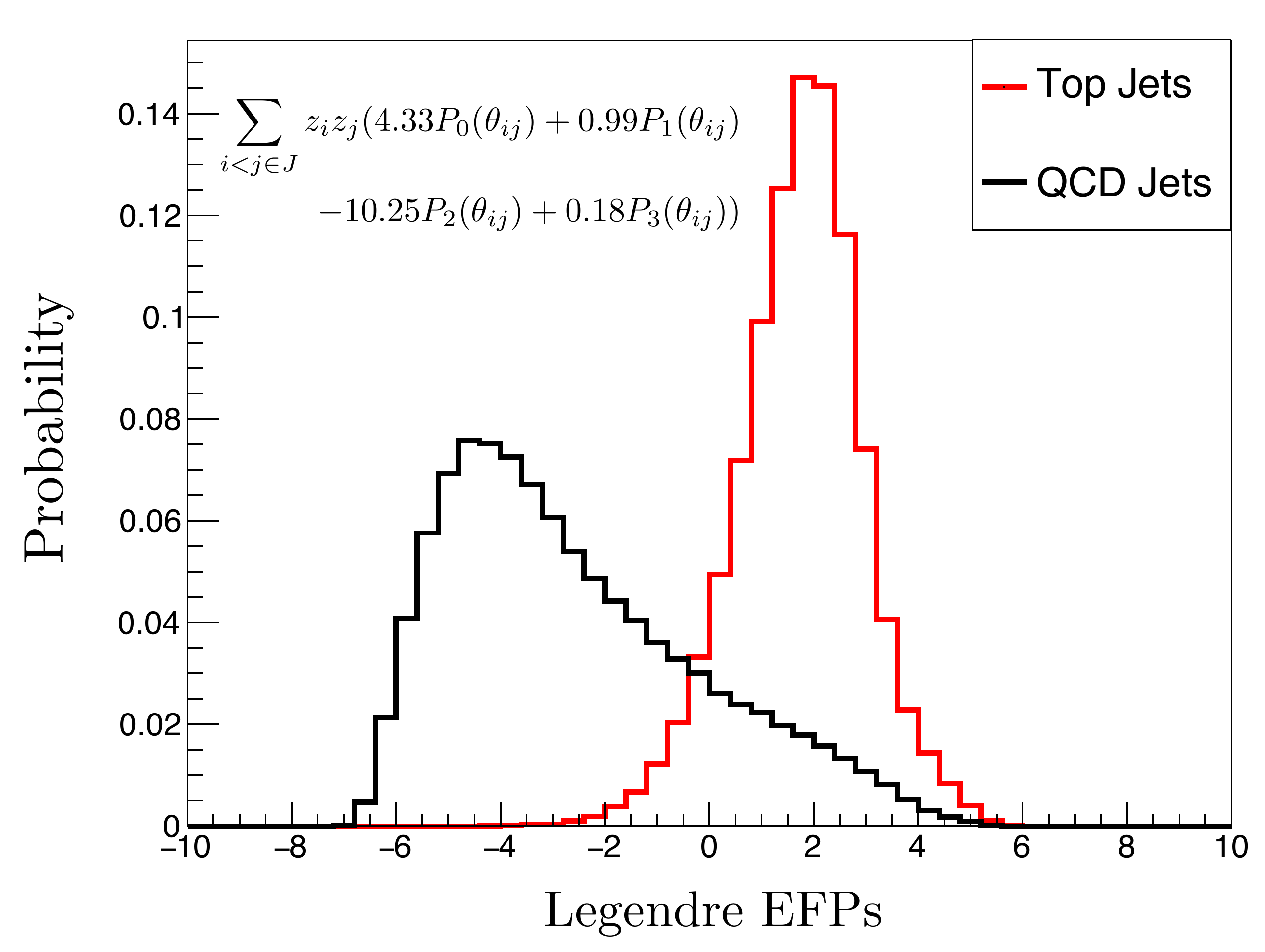}
\includegraphics[width=7.5cm]{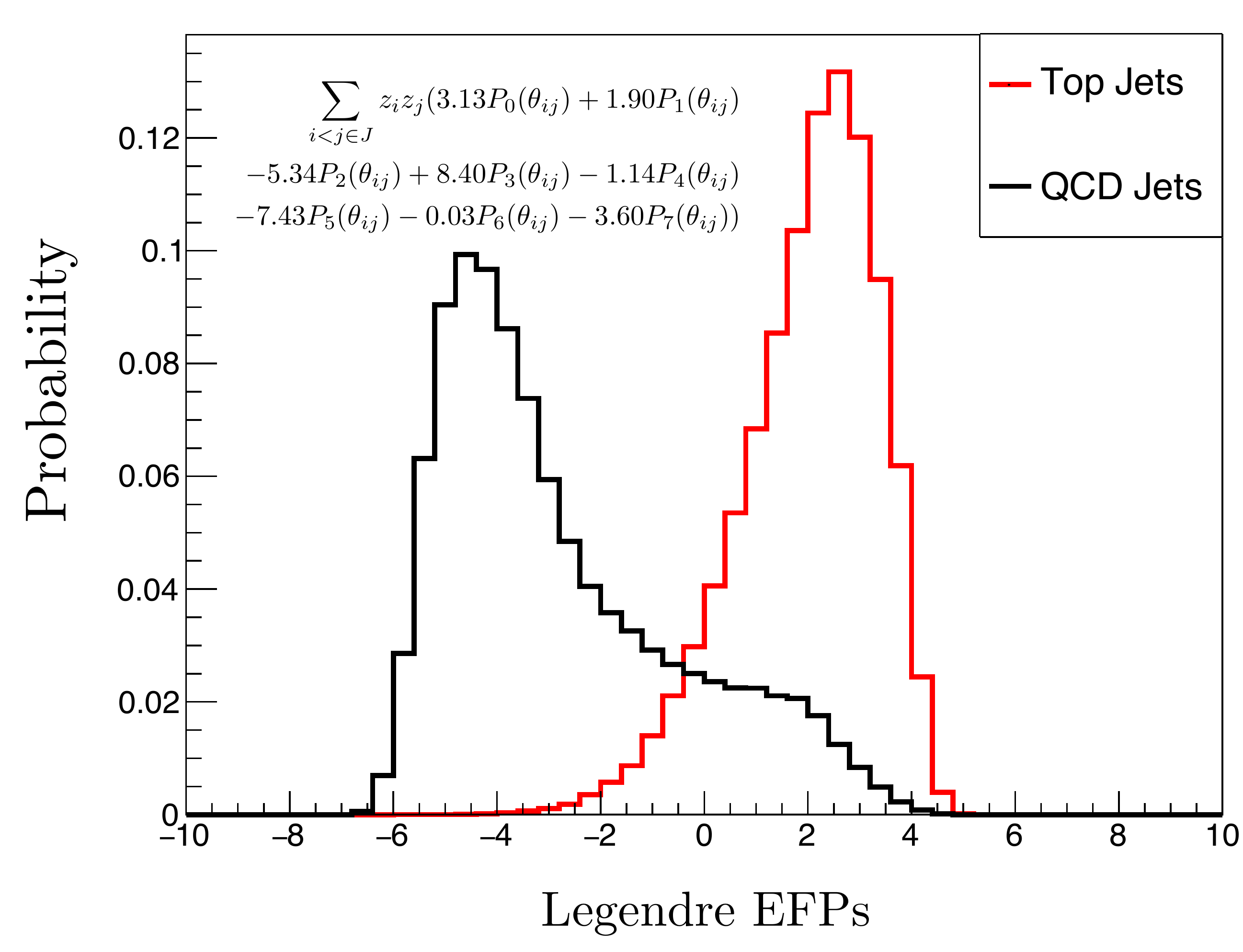}
\caption{Normalized distributions of the 2-point Legendre Energy Flow Polynomials for top jets and QCD jets that achieved the best classification performance. The corresponding truncation orders $\beta_{max}$ are 4 and 8, respectively. The specific form of the 2-point Legendre Energy Flow Polynomials is also indicated on the figure.}
\label{distributions2}
\end{center}
\end{figure*}

In Fig.\ref{distributions1}, we present the distributions of 4-point Legendre Hierarchical Energy Flow polynomials, denoted as \text{HEFP}(4, $\{\beta_{1},\beta_{2},\beta_{3}\}$), for both top jets and QCD jets, spanning across different parameter settings $(\{\beta_{1},\beta_{2},\beta_{3}\})$. Remarkably, the distributions of all the \text{HEFP}(4, ${\beta_{1},\beta_{2},\beta_{3}}$) of top jets and QCD jets exhibit clear discrimination. This distinct separation underscores the discriminative power of the \text{HEFP} approach and highlights its potential significance in distinguishing top jets from QCD jets in high energy physics analyses. Moreover, since $P_0(\theta_{ij})=1$,  \text{HEFP}(4, $\{0,\beta_{2},\beta_{3}\}$) will degenerate into 3-point \text{HEFP}(3, $\{\beta_{2},\beta_{3}\}$), while \text{HEFP}(4, $\{0,0,\beta_{3}\}$) will degenerate into 2-point \text{HEFP}(2, $\{\beta_{3}\}$). Therefore, the $(N+1)$-point HEFPs encompass all the HEFPs ranging from 2-point to $N$-point. As $\sum_i \beta_i$ increases, the HEFPs gradually move closer to the $y$-axis, indicating an increasing fraction of HEFPs that approach zero. This is primarily due to the cancellation effects observed in the higher order components. Consequently, for a specific truncation order $\beta_{max}$, any EFP observable with a path graph structure can be accurately expanded into a linear combination of $\beta_{max}^N$ HEFPs. For instance, we can easily employ the Linear Logistic Regression to identify the specific form of the 2-point \text{HEFP}(2, f) with the optimal classification performance. The number of undetermined parameters $\{\alpha_i\}$ is $\beta_{max}$, and we define the probability of the input jet being tagged as the top jet as
\begin{equation}
    p = \sigma(\sum_{\beta} \alpha_{\beta}\sum_{i,j}^{M}z_{i}z_{j}\text{P}_{\beta}(\theta_{ij})),
\end{equation}
where $\sigma(t)=1/(1+e^{-t})$ is the logical function. The objective function of logistic regression is 
\begin{equation}
    J(\alpha) = -[y \text{log}(p)+(1-y) \text{log}(1-p)],
\end{equation}
where $y=1$ corresponds to the top jet and $y=0$ corresponds to the QCD jet. 
In Fig.\ref{distributions2}, we present the normalized distributions of the 2-point Legendre Energy Flow Polynomials for top jets and QCD jets that achieved the optimal classification performance. The truncation orders $\beta_{max}$ for top jets and QCD jets are 4 and 8, respectively. Additionally, we illustrate the specific form of the 2-point Legendre Energy Flow Polynomials in this figure, which serves as a bridge between physical interpretability and machine learning. As shown in Fig.\ref{distributions2}, the distinctive form of the 2-point Legendre Energy Flow Polynomials, obtained through linear regression, proves to be highly effective in discriminating between top jets and QCD jets. 

Furthermore, we find that the above HEFP(N, $\text{P}_{\beta}$) obtained under a fixed energy-weighted summation order exhibits IRC-safety after each node elimination. This special characteristic, which we called hierarchical IRC-safety, indicates that inserting arbitrary functions (including non-linear function) after each energy-weighted summation is allowed, where the parameterized and trainable neural networks can come into and play an important role. For example, 
\begin{equation}
    \mathcal{O}^{(2)} = \sum_{i}^{M}z_{i}\Phi(\sum_{j}^{M}z_{j}\text{P}_{\beta}(\theta_{ij}))
\end{equation}
where 2 indicates that the features are 2-point energy flow observable. After performing variable elimination algorithm, any EFPs can be computed via a hierarchical summation like
\small
\begin{eqnarray*}
    && \sum_{i,j,k,l}^{M}z_{i}z_{j}z_{k}z_{l}f_{1}(\theta_{ij})f_{2}(\theta_{jk})f_{3}(\theta_{jl})f_{4}(\theta_{kl})\nonumber \\
        && =\sum_{\beta_{1},\beta_{2},\beta_{3},\beta_{4}}^{\beta_{max}}\alpha_{\beta_{1}}\alpha_{\beta_{2}}\alpha_{\beta_{3}}\alpha_{\beta_{4}}(\sum_{i}^{M}z_{i}(\sum_{j}^{M}z_{j}\text{P}_{\beta_{1}}(\theta_{ij})(\sum_{k}^{M}z_{k}\text{P}_{\beta_{2}}(\theta_{jk})(\sum_{l}^{M}z_{l}\text{P}_{\beta_{3}}(\theta_{kl})\text{P}_{\beta_{4}}(\theta_{jl}))))),
\end{eqnarray*}
\normalsize 
where $\alpha_{\beta}$ are linear coefficients of orthogonal bases with different orders. Moreover, an arbitrary function $\Phi$ can be inserted IRC-safely between each energy-weighted summation and the next layer calculation, which can be effectively parameterized and trained by using neural networks. In the proceeding section, we will introduce a novel backbone model that utilizes neural network parameterization to reconstruct the path-graph energy flow polynomials while simultaneously maintaining interpretability, IRC-safety, and rotational invariance. In this work, we only consider the path-graph corresponding EFPs. For simplicity, all the Hierarchical Energy Flow functions specifically refer to path-graph Hierarchical Energy flow functions in the following sections. However, it is essential to highlight that this framework actually can be generalized to any other complex graph structures.

\section{Hierarchical Higher-Point Energy Flow Network}
\subsection{Hierarchical Energy Flow Network}
As mentioned in the preceding section, inserting an arbitrary function after the energy-weighted summation of Hierarchical Energy Flow Polynomials does not break the IRC safety. Thus we can parameterize these functions with neural networks. For instance, in the case of 2-point HEFN, the 1-point features of $i$-th particle $\hat{p}_{i}^{1}$ is embedded initially as 
\begin{equation}
    \hat{p}_{i}^{1} = \Phi(\sum_{j}^{M}z_{j}\text{P}_{\beta}(\theta_{ij})),
\end{equation}
where $\hat{p}_{i}^{1}$ are invariant under translation, rotation, or reflection of the jet in the angular direction. Consequently, the jet observables in the latent space are also IRC-safe and can be expressed as
\begin{equation}
    \mathcal{O}^{(2)} = F(\sum_{i}z_{i}\Phi(\sum_{j}^{M}z_{j}\text{P}_{\beta}(\theta_{ij}))),
\end{equation}
which are natural generalizations of EFNs when considering 2-point energy flow functions. We introduce two MLP modules $\Phi$ and $F$ as EFNs do. The first MLP $\Phi:\mathcal{R}^{\beta_{max}}\rightarrow \mathcal{R}^{l}$ maps 1-point particle features with dimension $\beta_{max}$, the truncated order of the Legendre polynomials, into a latent space with dimension $l$, while the 2-point jet observables are energy-weighted summation of all particle features. The second MLP can be viewed as a discriminant of jet features. We find better tagging performance in some public datasets compared with EFNs, as discussed in Section 4.

In the following steps, we extend the scenario to $N$-point HEFN. 
We incorporate neural networks into Hierarchical Energy Flow Polynomials to obtain Hierarchical Energy Flow Networks (HEFNs). For $t$-point particle features at latent space $\hat{p}_{i}^{t}$, the next higher-point particle features are obtained through recursive relations as 
\begin{equation}
    \hat{p}_{i}^{t+1} = \Phi^{a}(\sum_{j}^{M}z_{j}\Phi^{b}(\hat{p}_{j}^{t})\otimes \text{P}_{\beta}(\theta_{ij}))+\hat{p}_{i}^{t}.
\label{recursive}
\end{equation}
The second MLP $\Phi^{b}:\mathcal{R}^{l}\rightarrow \mathcal{R}^{d}$ maps $t$-point particle features $\hat{p}^{t}_{j}$ into dimension-$d$. And we fix the dimension of hidden space $l^{\prime} = d\times \beta_{max}$ to control model parameter capability for different $\beta_{max}$ setting. The first MLP $\Phi^{a}:\mathcal{R}^{l^{\prime}}\rightarrow \mathcal{R}^{l}$ maps the hidden $(t+1)$-point particle features back to latent space. Besides, the residual connection we introduce here makes
\begin{equation}
    \hat{p}^{t+1}_{i} = \sum_{k=1}^{t}\Phi^{a,k}(\sum_{j}^{M}z_{j}\Phi^{b,k}(\hat{p}_{j}^{k})\otimes \text{P}_{\beta}(\theta_{ij})),
\label{resnet}
\end{equation}
which could let the final jet observables contain a mix of multiple high-point energy flow functions 
\begin{equation}
    \mathcal{O}^{(N+1)}=F(\sum_{i}^{M}z_{i}\sum_{k=1}^{N}\Phi^{a,k}(\sum_{j}^{M}z_{j}\Phi^{b,k}(\hat{p}_{j}^{k})\otimes \text{P}_{\beta}(\theta_{ij}))). 
\end{equation}
It is worth noting that since the inserted nonlinear functions are not only about angular information, the learned observables cannot be expanded as a linear combination of C-correlators nor EFPs. 
In the generalized scenario of an $N$-point Hierarchical Energy Flow Network before the final discriminant F, the architecture consists of total $N$ $\Phi^{a}$ and $(N-1)$ $\Phi^{b}$ since we only apply $\Phi^{a}$ to get $\hat{p}_{i}^{1}$. Both $\Phi^{a}$ and $\Phi^{b}$ are MLPs with BatchNorm-ReLU-FullyConnected structures. 
Note that when applying Batch Normalization to particle-level features, we start by obtaining jet-level observables through energy-weighted summation. Subsequently, we calculate the mean and variance of jet features in mini-batches. This ensures that Batch Normalization maintains the statistical robustness of neural network inputs, even under particle splitting and soft radiation conditions. 

In our study, we keep the dimension of particle features at latent space $l=256$ and hidden space $l^{\prime}=d\times \beta_{max}=1024$. The first MLP $\Phi^{a}$ is stacked with two BN-ReLU-FC blocks with (1024, 256, 256) nodes, while the second MLP $\Phi^{b}$ is a single BN-ReLU-FC block with (256, d) nodes. We selected $\beta_{max}$ values of 4, 8, 16, and 32 as the truncated orders of the Legendre polynomial to investigate their impact on the model's performance. Additionally, we systematically compared the performance of up to 2, 3, and 4-point HEFN (Hierarchically Energy Flow Network) on top tagging dataset and quark/gluon discrimination dataset. All the results will be presented in Section \ref{Results of Jet Classification}.

\subsection{Local Hierarchical Energy Flow Network.}

To further reduce the computational complexity of the model while preserving essential information, we ignore the contribution of distant particles and adopt the neighbour aggregation approach instead of global summation. In other words, we require $f(R_{ij})=0$ if $R_{ij}>r_{max}$. Besides, it is also possible to pixelate the distance between particles to aggregate particles in different regions~\cite{pointnet++}. Based on the neighbour aggregation, we propose the second strategy, the Local Hierarchical Energy Flow Network (LHEFN).

In our case, we choose to use the query ball approach to define neighborhood for IRC-safety during end-to-end training~\cite{ewmp}. To combine features from different scales, we group multi-scale neighborhoods to construct the graph structure as 
 \begin{equation}
     A_{ij}^{k}=\left\{
     \begin{aligned}
     1 & \quad \text{if} \quad (k-1)\epsilon \leq R_{ij}<k\epsilon\\
     0 & \quad \text{else} \\ 
     \end{aligned}
     \right.
 \end{equation}
where $\epsilon$ measures the resolution of our implementation.
The value of $r_{max}=k_{max}\epsilon$ represents the maximum scope radius.

We initialize 1-point particle features similar with HEFN:
\begin{equation}
    \hat{p}_{i}^{1} = \Phi(\sum_{j}z_{j}A_{ij}^{k}).
\end{equation}
Currently, we continue to iteratively update the particle features following Eq. (\ref{recursive}). However, instead of relying on pair-wise features computed using Legendre polynomials, we now adopt $A_{ij}^{k}$ to replace them. Note that $A_{ij}^{k}$ are symmetric edge interaction tensors resulting in a radial single square wave form. 
Since the edge embeddings are binary functions, the nodes can be updated using a simple weighted summation operator, significantly reducing computational complexity. The update step of $t$-point particle features can be expressed as
\begin{equation}
    \hat{p}^{t+1}_{i}=\Phi([\hat{p}^{t}_{i,1}, \hat{p}^{t}_{i, 2}, ... , \hat{p}^{t}_{i, k_{max}}]),
    \label{update}
\end{equation}
where $\hat{p}^{t}_{i,k}=\sum_{j}^{A_{ij}^k=1} z_j \hat{p}^{t}_{j}$ is the energy-weighted sum of all particle features between distances k$\epsilon$ and (k+1)$\epsilon$ away from particle $i$. This computation efficiently aggregates relevant information from neighboring particles, taking into account their energy contributions. The neural network parameterized function $\Phi$ plays a crucial role in adaptively selecting the aggregation radius, and combines features of particles on different neighbor regions. Note that this approach may exhibit limited flexibility in expressing edge information, 
resulting in potentially worse performance compared to the Legendre polynomial encoding. However, the aforementioned formula significantly reduces the computational complexity of $O(d\cdot M^2 \cdot \beta$) compared with Eq. (\ref{resnet}) by eliminating the need for the direct product of edge features and particle features. The model gains a remarkable boost in speed, making it more practical for real applications. 
To ensure fair evaluations, we reset $r_{max}=0.5R_0,\; R_0,\; 1.5R_0\; \text{and}\; 2R_0$ and selected $k_{max}$ of 4, 8, 16, and 32 for testing purposes. Additionally, we maintained consistency in the remaining trainable parameter capacity and hyperparameter settings throughout the experiments.

\subsection{Model Implementation}
The model architecture is implemented in the PYTORCH deep learning framework with the CUDA platform. We adopt the binary cross-entropy as the loss function. To optimize the model parameters, we employ the Adam optimizer~\cite{Kingma:2014vow} with an initial learning rate of 0.001 and momentum set to 0.8, which is determined based on the gradients calculated on a mini-batch of 128 training examples. The network is trained up to 50 epochs, with a cosine decay learning rate scheduler. In addition, we employ the early-stopping technique to prevent over-fitting.

\section{Results of Jet Classification}
\label{Results of Jet Classification}

\subsection{Top Tagging}
We perform the top tagging analysis utilizing a benchmark dataset~\cite{Benato:2021olt}, containing hadronic tops as the signal and QCD di-jets as the background. The event generation is performed by Pythia8~\cite{Sjostrand:2014zea}, while the detector effect is simulated by Delphes~\cite{deFavereau:2013fsa}. The particle-flow constituents are clustered into jets using the anti-kT algorithm~\cite{Cacciari:2008gp} with $R_0$ = 0.8 as the radius parameter. Our analysis focuses on jets with transverse momentum $p_T \in$ [550, 650] GeV and rapidity $|y| < 2$. The dataset consists of 1.2M training events, 400k validation events, and 400k test events. Only the energy-momentum 4-vectors for each particle inside the jets are contained.  Both the HEFN architecture and the LHEFN architecture utilize only $p_T$, $\eta$, and $\phi$ of all constituent particles inside each jet. To ensure dataset quality, jets with a particle count of less than 5 were removed to prevent the influence of extreme examples on the datasets (0.01$\%$ for the top tagging dataset).

\begin{figure}[ht]
    \centering
    \includegraphics[width=16cm,height=16cm]{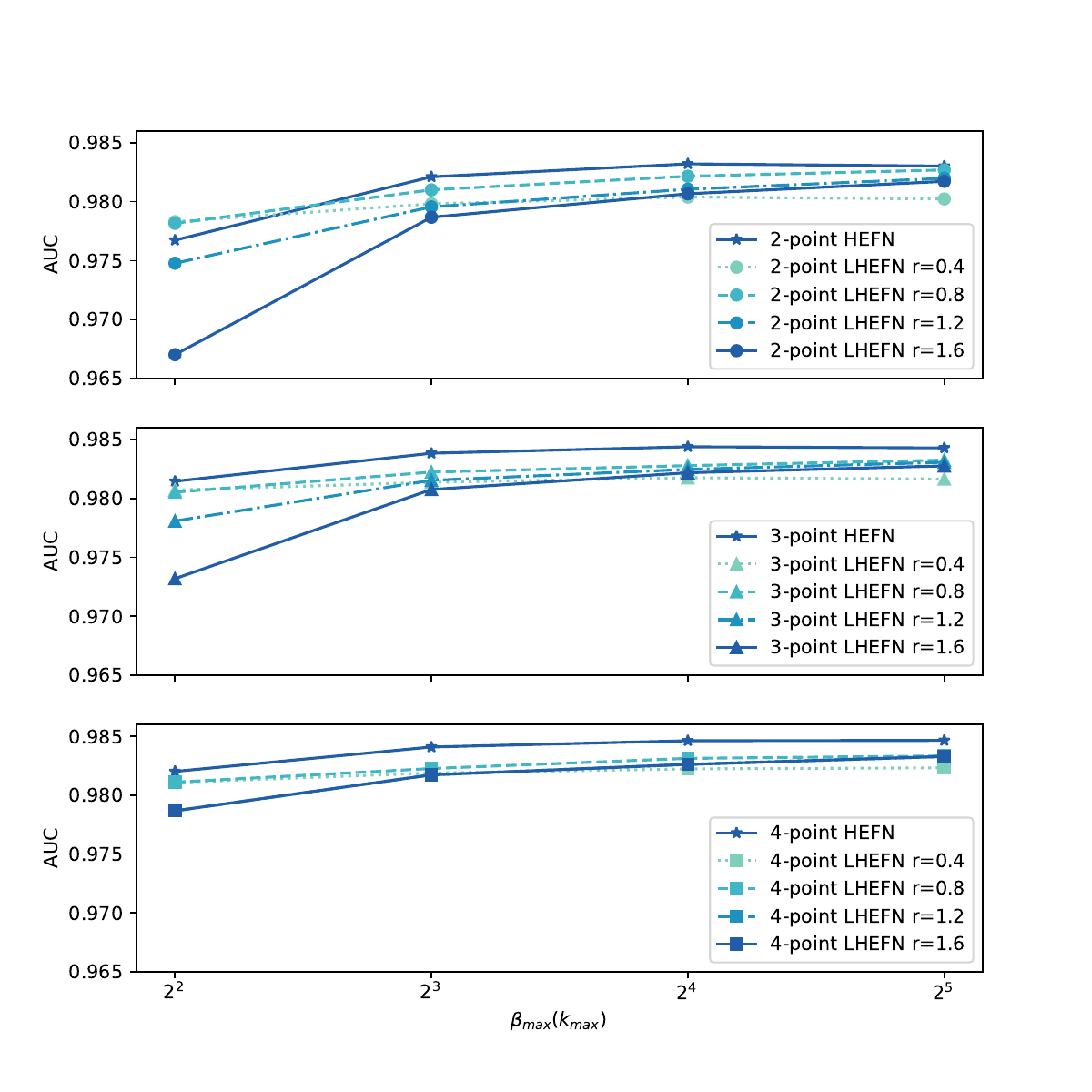}
    \vspace*{-1.5cm}
    \caption{Area under the ROC curve (AUC) for top tagging task of 2-point, 3-point, and 4-point HEFN as a function of $\beta_{max}$, and AUC for top tagging task of 2-point, 3-point, and 4-point LHEFN with varying $r$ values ($r = 0.4, 0.8, 1.2, 1.6$) as a function of $k_{max}$, where $r$ represents the maximum scope radius for LHEFN. The $x$-axis is shared for both $\beta_{max}$ and $k_{max}$.}
    \label{top}
\end{figure}

Fig.\ref{top} shows the area under the ROC curve (AUC) for the top tagging task of 2-point, 3-point, and 4-point HEFN as a function of $\beta_{max}$, along with the AUC for the top tagging task of 2-point, 3-point, and 4-point LHEFN with varying $r$ values ($r = 0.4, 0.8, 1.2, 1.6$) as a function of $k_{max}$. The results displayed in Fig.\ref{top} exhibit a notable trend where the Area Under the Curve (AUC) of HEFN consistently increases with $N$. This observation indicates that increasing the complexity of the network by adding more points (2-point, 3-point, and 4-point configurations) can enhance the performance of HEFN. Additionally, the AUC remains almost unchanged during the transition from 3-point to 4-point configurations, which  indicates that the 4-point HEFN achieves the optimal performance. For all the considered 2-point, 3-point, and 4-point HEFN architectures, we notice a consistent improvement in AUC as we increase the truncated orders of the Legendre Polynomials $\beta_{max}$. Similar behavior is observed in the case of 2-point, 3-point, and 4-point LHEFN models with varying $r$ values (0.4, 0.8, 1.2, and 1.6). As the parameter $k_{max}$ increases, the AUC of LHEFN also demonstrates a notable enhancement, signifying that refining the neighbor division leads to a higher AUC. Another noteworthy discovery is that setting the value of $r$ empirically to half the radius of the candidate jets yields optimal performance for LHEFN. Although it is worth noting that HEFN outperforms LHEFN slightly across almost all parameter settings, the LHEFN is notably faster compared to HEFN.  The choice between the two models may depend on specific application requirements, with HEFN offering higher AUC and LHEFN presenting a notable speed advantage.

In Table \ref{tab:results_top} we provide detailed results of accuracy, AUC, and background rejection for the optimal parameter settings ($N=4$, $\beta_{max}=16$) and ($N=4$, $k_{max}=16$) of both HEFN and LHEFN. Additionally, we present the performance achieved by various classification algorithms on the top tagging dataset, facilitating a comprehensive comparison. From Table \ref{tab:results_top}, it is evident that by solely utilizing the $p_T$, $\eta$, and $\phi$ information of all constituent particles, both HEFN and LHEFN deliver comparable performance to existing model architectures. Moreover, both HEFN and LHEFN enable the reconstruction of energy correlation-based observables while maintaining interpretability, IRC-safety, and rotational invariance, which other models can not preserve.

\begin{table}[htb]
    \centering
    \caption{Performance comparison of HEFN and LHEFN with Existing Classification Algorithms on the Top Tagging Dataset. The uncertainty is calculated by taking the standard deviation of 5 training runs with different random weight initialization.}
    \label{tab:results_top}
	\begin{tabular}{lccccc}
          &  Accuracy &AUC & 1/$\epsilon_B$ ($\epsilon_S = 0.5$)  & 1/$\epsilon_B$ ($\epsilon_S = 0.3$) \\
            \hline
            ResNeXt-50 \cite{Qu:2019gqs} & 0.936 & 0.9837 & 302$\pm$5 & 1147$\pm$58 \\
            P-CNN \cite{Qu:2019gqs} & 0.930 & 0.9803 & 201$\pm$4 & 759$\pm$24 \\
            PFN \cite{Komiske:2018cqr} & - & 0.9819 & 247$\pm$3 & 888$\pm$17 \\
            ParticleNet-Lite \cite{Qu:2019gqs} & 0.937 & 0.9844 &  325$\pm$5 & 1262$\pm$49 \\
            ParticleNet \cite{Qu:2019gqs} & 0.940 & 0.9858 & 397$\pm$7 & 1615$\pm$93\\
            JEDI-net  \cite{Moreno:2019bmu}  & 0.9263 & 0.9786 & - & 590.4 \\
            JEDI-net with $\sum O$  \cite{Moreno:2019bmu}  & 0.9300 & 0.9807 & - & 774.6 \\
            SPCT \cite{Mikuni:2021pou} & 0.928 & 0.9799 & 201$\pm$9 & 725$\pm$54 \\
            PCT \cite{Mikuni:2021pou} & 0.940 & 0.9855 & 392$\pm$7 & 1533$\pm$101 \\
            LorentzNet \cite{Gong:2022lye} & 0.942 & 0.9868 & 498$\pm$18 & 2195$\pm$173 \\
            ParT \cite{Qu:2022mxj} & 0.940 & 0.9858 & 413$\pm$16 & 1602$\pm$81 \\
            \hline
            \textbf{HEFN} & \textbf{0.9375} & \textbf{0.9846} & \textbf{343$\pm$6} & \textbf{1262$\pm$51} \\
            \textbf{LHEFN} & \textbf{0.9337} & \textbf{0.9833} & \textbf{271$\pm$5} & \textbf{935$\pm$21} \\
	\end{tabular}
\end{table}

\subsection{Quark/Gluon Discrimination}

The Quark-Gluon benchmark dataset~\cite{Komiske:2018cqr}, generated using Pythia8 without detector simulation, consists of quark-initiated samples $q\overline{q} \rightarrow{Z\rightarrow{\nu\overline{\nu}}+(u,d,s)}$ as the signal and gluon-initiated data $q\overline{q} \rightarrow{Z\rightarrow{\nu\overline{\nu}}+g}$ as the background. For jet clustering, we use the anti-kT algorithm with $R_0$ = 0.4. Our analysis focuses on selecting jets with transverse momentum $p_T \in$ [500, 550] GeV and rapidity $|y| < 1.7$. Each particle in the dataset is characterized by its four-momentum and particle identification (PID) information. However, in this study, both the HEFN and LHEFN architectures utilize only $p_T$, $\eta$, and $\phi$ of all the constituent particles inside each jet. The official dataset comprises of a total of 2 million events. Among these, 1.6 million events are used for training, while 200k events each are allocated for validation and testing. Besides, to ensure dataset quality, jets with a particle count of less than 5 are removed to avoid any adverse influence of extreme examples on the datasets. The proportion of these filtered samples relative to the entire dataset is minimal, representing only 0.01$\%$ of the Quark/Gluon Discrimination dataset.

\begin{figure}[ht]
    \centering
    \includegraphics[width=16cm,height=16cm]{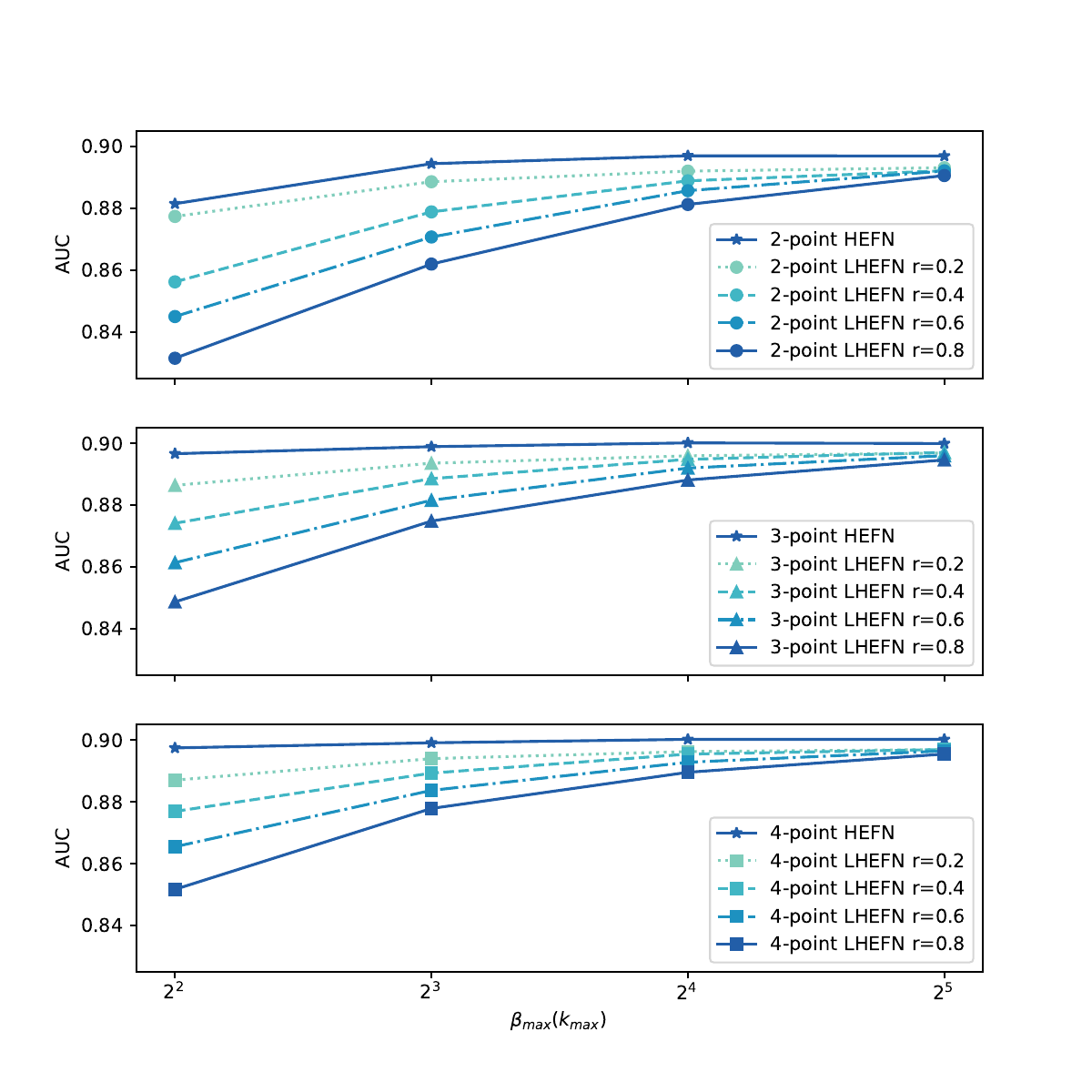}
     \vspace*{-1.5cm}
    \caption{Area under the ROC curve (AUC) for quark/gluon discrimination task of 2-point, 3-point, and 4-point HEFN as a function of $\beta_{max}$, and AUC for top tagging task of 2-point, 3-point, and 4-point LHEFN with varying $r$ values ($r = 0.2, 0.4, 0.6, 0.8$) as a function of $k_{max}$, where $r$ represents the maximum scope radius for LHEFN. The $x$-axis is shared for both $\beta_{max}$ and $k_{max}$.}
    \label{qg}
\end{figure}

Fig.\ref{qg} shows the area under the ROC curve (AUC) for the quark/gluon discrimination task of 2-point, 3-point, and 4-point HEFN as a function of $\beta_{max}$, along with the AUC for the quark/gluon discrimination task of 2-point, 3-point, and 4-point LHEFN with $r = 0.2, 0.4, 0.6, 0.8$ as a function of $k_{max}$, where $r$ is the maximum scope radius of LHEFN. The results presented in Fig.\ref{qg} show a clear pattern: the Area Under the Curve (AUC) of HEFN consistently increases with $N$, which suggests that the performance of HEFN benefits from the increased complexity brought by adding more points. Moreover, during the shift from 3-point to 4-point configurations, the AUC remains relatively stable, implying that the 4-point HEFN reaches its optimal performance at this stage. Regarding the 2-point HEFN architecture, we observe that increasing the truncated orders of the Legendre Polynomials $\beta_{max}$ leads to a consistent improvement in the Area Under the Curve (AUC). However, for the 3-point and 4-point HEFN architectures, the AUC remains unchanged as $\beta_{max}$ increases. Different from HEFN, in the case of 2-point, 3-point, and 4-point LHEFN models with $r = 0.2, 0.4, 0.6, 0.8$, as the parameter $k_{max}$ increases, there is a notable enhancement in the AUC of LHEFN, indicating that refining the neighbor division can result in higher AUC. Another noteworthy finding is that setting the value of $r_{max}$ to half the radius of the candidate jets results in optimal performance for LHEFN. Although it is worth noting that HEFN performs slightly better than LHEFN across all parameter settings, LHEFN stands out for its significantly faster performance compared to HEFN. As a result, the selection between the two models may depend on the specific needs of the application, where HEFN is preferred for achieving higher AUC, while LHEFN is a more suitable choice when speed is a critical factor.

In Table \ref{tab:results_qg}, we present the results of accuracy, area under the curve (AUC), and background rejection for the parameter settings ($N=4$, $\beta_{max}=$16) and ($N=4$, $k_{max}=$16) of HEFN and LHEFN which correspond to the optimal performance. Additionally, we present the performance attained by several classification algorithms on the quark and gluon dataset, enabling a comprehensive comparison. From Table \ref{tab:results_top}, we can see clearly that both HEFN and LHEFN can achieve competitive performance compared with the existing architectures even with only the $p_T$, $\eta$, and $\phi$ information of all constituent particles. More importantly, both HEFN and LHEFN can reconstruct energy correlation-based observables, setting them apart from other models that fail to preserve crucial features including interpretability, IRC-safety, and rotational invariance.

\begin{table}[htb]
    \centering
    \caption{Performance comparison of HEFN and LHEFN with Existing Classification Algorithms on the Quark/Gluon Discrimination Dataset. The uncertainty is calculated by taking the standard deviation of 5 training runs with different random weight initialization. Note that in the HEFN and LHEFN networks, we did not utilize any PID information.}
    \label{tab:results_qg}
	\begin{tabular}{lccccc}
          &  Accuracy &AUC & 1/$\epsilon_B$ ($\epsilon_S = 0.5$)  & 1/$\epsilon_B$ ($\epsilon_S = 0.3$) \\
            \hline
            ResNeXt-50 \cite{Qu:2019gqs} & 0.821 & 0.9060 & 30.9 & 80.8 \\
            P-CNN \cite{Qu:2019gqs} & 0.827 & 0.9002 & 34.7 & 91.0 \\
            PFN \cite{Komiske:2018cqr} & - & 0.9005 & 34.7$\pm$0.4 & - \\
            ParticleNet-Lite \cite{Qu:2019gqs} & 0.835 & 0.9079 &  37.1 & 94.5 \\
            ParticleNet \cite{Qu:2019gqs} & 0.840 & 0.9116 & 39.8$\pm$0.2 & 98.6$\pm$1.3\\
            ABCNet \cite{Mikuni:2020wpr}& 0.840 & 0.9126 & 42.6$\pm$0.4 & 118.4$\pm$1.5 \\
            SPCT \cite{Mikuni:2021pou} & 0.815 & 0.8910 & 31.6$\pm$0.3 & 93.0$\pm$1.2 \\
            PCT \cite{Mikuni:2021pou} & 0.841 & 0.9140 & 43.2$\pm$0.7 & 118.0$\pm$2.2 \\
            LorentzNet \cite{Gong:2022lye} & 0.844 & 0.9156 & 42.4$\pm$0.4 & 110.2$\pm$1.3 \\
            ParT \cite{Qu:2022mxj} & 0.849 & 0.9203 & 47.9$\pm$0.5 & 129.5$\pm$0.9 \\
            \hline
            \textbf{HEFN} & \textbf{0.8264} & \textbf{0.9002} & \textbf{33.5$\pm$0.3} & \textbf{86.2$\pm$0.8}\\
            \textbf{LHEFN} & \textbf{0.8213} & \textbf{0.8969} & \textbf{31.3$\pm$0.2} & \textbf{82.5$\pm$0.7} \\

 \end{tabular}
\end{table}

\section{Conclusions}

In this work, we introduced a novel IRC-safe deep learning framework for analyzing high-point energy flow observables in Jet Substructure analysis. We utilized a class of high-point energy flow functions from Energy Flow Polynomials as a comprehensive basis for observables, enabling hierarchical energy-weighted summation. By incorporating orthogonal polynomials to quantify the correlation between particles based on angular distance and employing neural networks as parametrized functions, we achieved enhanced flexibility in representation without compromising IRC safety.

Our approach demonstrated both interpretability and remarkable discrimination performance in top tagging dataset and quark-gluon dataset. To strike a balance between computational complexity and model performance, we efficiently reduced global summation to local aggregation and encoded multi-scale neighborhoods with one-hot encoding. This enabled the creation of Local Hierarchical Energy Flow Networks (LHEFN) based on local energy flow, showing comparable performance to the HEFN.

Overall, our IRC-safe deep learning framework can provide a simple yet powerful tool for analyzing high-point energy flow observables in Jet Substructure analysis, with potential applications in various high-energy collision scenarios. The combination of comprehensive observables, orthogonal polynomials, and neural networks allows us to achieve excellent performance while maintaining IRC safety, rotation invariance and interpretability, making it a valuable addition to the field of jet substructure analysis.

Furthermore, there are several potential improvements that could be explored in future research. Firstly, in this study we relaxed the Lorentz invariance to SO(2)-rotational symmetry around the jet axis. To ensure strict preservation of Lorentz symmetry, we could consider replacing the angular distance $\theta_{ij}$ with a Lorentz invariant quantity, such as invariant mass, within the Hierarchical Energy Flow Polynomials parametrization. Secondly, for the quark/gluon discrimination task, we did not utilize the Particle Identification (PID) information. Incorporating the PID information into the model architecture could potentially lead to performance improvements, as it provides valuable additional information for distinguishing gluon jet from light quark jet. Thirdly, in our approach, we solely reconstructed the Energy Flow Polynomials observables with path graph structures. It would be interesting to explore more complex graph structures and investigate the reconstruction of EFP observables corresponding to different graph configurations, as this could offer further insights into the energy distribution within the jets. These possible improvements merit further investigation in future studies. By addressing these aspects, we can enhance the capabilities of our model and potentially uncover new insights into jet substructure analysis.

\addcontentsline{toc}{section}{Acknowledgments}
\acknowledgments
We thank Lei Wu for valuable discussions. This work was supported by the National Natural Science Foundation of China
(NNSFC) under grant Nos.11821505 and 12075300,
by Peng-Huan-Wu Theoretical Physics Innovation Center (12047503),
by the CAS Center for Excellence in Particle Physics (CCEPP),
and by the Key Research Program of the Chinese Academy of Sciences, Grant NO. XDPB15.

\addcontentsline{toc}{section}{References}
\bibliographystyle{JHEP}
\bibliography{bibliography}

\providecommand{\href}[2]{#2}\begingroup\raggedright\begin{thebibliography}{10}

\bibitem{whatisjet}
F.~V. Tkachov, \emph{Measuring multijet structure of hadronic energy flow or,
  what is a jet?},
  \href{https://doi.org/10.1142/s0217751x97002899}{\emph{International Journal
  of Modern Physics A} {\bfseries 12} (dec, 1997) 5411--5529}.

\bibitem{gurari2011classification}
G.~Gur-Ari, M.~Papucci and G.~Perez, \emph{Classification of energy flow
  observables in narrow jets},  2011.

\bibitem{1011.2268}
J.~Thaler and K.~V. Tilburg, \emph{Identifying boosted objects with
  n-subjettiness}, \href{https://doi.org/10.1007/jhep03(2011)015}{\emph{Journal
  of High Energy Physics} {\bfseries 2011} (mar, 2011) }.

\bibitem{1108.2701}
J.~Thaler and K.~V. Tilburg, \emph{Maximizing boosted top identification by
  minimizing n-subjettiness},
  \href{https://doi.org/10.1007/jhep02(2012)093}{\emph{Journal of High Energy
  Physics} {\bfseries 2012} (feb, 2012) }.

\bibitem{1305.0007}
A.~J. Larkoski, G.~P. Salam and J.~Thaler, \emph{Energy correlation functions
  for jet substructure},
  \href{https://doi.org/10.1007/jhep06(2013)108}{\emph{Journal of High Energy
  Physics} {\bfseries 2013} (jun, 2013) }.

\bibitem{1609.07483}
I.~Moult, L.~Necib and J.~Thaler, \emph{New angles on energy correlation
  functions}, \href{https://doi.org/10.1007/jhep12(2016)153}{\emph{Journal of
  High Energy Physics} {\bfseries 2016} (dec, 2016) }.

\bibitem{efps}
P.~T. Komiske, E.~M. Metodiev and J.~Thaler, \emph{Energy flow polynomials: a
  complete linear basis for jet substructure},
  \href{https://doi.org/10.1007/jhep04(2018)013}{\emph{Journal of High Energy
  Physics} {\bfseries 2018} (apr, 2018) }.

\bibitem{Marzani_2019}
S.~Marzani, G.~Soyez and M.~Spannowsky, \emph{Looking Inside Jets}.
\newblock Springer International Publishing, 2019,
  \href{https://doi.org/10.1007/978-3-030-15709-8}{10.1007/978-3-030-15709-8}.

\bibitem{Berger_2004}
C.~Berger and L.~Magnea, \emph{Scaling of power corrections for angularities
  from dressed gluon exponentiation},
  \href{https://doi.org/10.1103/physrevd.70.094010}{\emph{Physical Review D}
  {\bfseries 70} (nov, 2004) }.

\bibitem{Almeida_2009}
L.~G. Almeida, S.~J. Lee, G.~Perez, G.~Sterman, I.~Sung and J.~Virzi,
  \emph{Substructure of high-pt jets at the {LHC}},
  \href{https://doi.org/10.1103/physrevd.79.074017}{\emph{Physical Review D}
  {\bfseries 79} (apr, 2009) }.

\bibitem{Albertsson:2018maf}
K.~Albertsson et~al., \emph{{Machine Learning in High Energy Physics Community
  White Paper}},
  \href{https://doi.org/10.1088/1742-6596/1085/2/022008}{\emph{J. Phys. Conf.
  Ser.} {\bfseries 1085} (2018) 022008},
  [\href{https://arxiv.org/abs/1807.02876}{{\ttfamily 1807.02876}}].

\bibitem{Abdughani:2019wuv}
M.~Abdughani, J.~Ren, L.~Wu, J.~M. Yang and J.~Zhao, \emph{{Supervised deep
  learning in high energy phenomenology: a mini review}},
  \href{https://doi.org/10.1088/0253-6102/71/8/955}{\emph{Commun. Theor. Phys.}
  {\bfseries 71} (2019) 955},
  [\href{https://arxiv.org/abs/1905.06047}{{\ttfamily 1905.06047}}].

\bibitem{Plehn:2022ftl}
T.~Plehn, A.~Butter, B.~Dillon and C.~Krause, \emph{{Modern Machine Learning
  for LHC Physicists}},  \href{https://arxiv.org/abs/2211.01421}{{\ttfamily
  2211.01421}}.

\bibitem{Cheng:2022idp}
T.~Cheng, \emph{{Bridging Machine Learning and Sciences: Opportunities and
  Challenges}},  \href{https://arxiv.org/abs/2210.13441}{{\ttfamily
  2210.13441}}.

\bibitem{Roe_2005}
B.~P. Roe, H.-J. Yang, J.~Zhu, Y.~Liu, I.~Stancu and G.~McGregor, \emph{Boosted
  decision trees as an alternative to artificial neural networks for particle
  identification},
  \href{https://doi.org/10.1016/j.nima.2004.12.018}{\emph{Nuclear Instruments
  and Methods in Physics Research Section A: Accelerators, Spectrometers,
  Detectors and Associated Equipment} {\bfseries 543} (may, 2005) 577--584}.

\bibitem{Larkoski:2017jix}
A.~J. Larkoski, I.~Moult and B.~Nachman, \emph{{Jet Substructure at the Large
  Hadron Collider: A Review of Recent Advances in Theory and Machine
  Learning}}, \href{https://doi.org/10.1016/j.physrep.2019.11.001}{\emph{Phys.
  Rept.} {\bfseries 841} (2020) 1--63},
  [\href{https://arxiv.org/abs/1709.04464}{{\ttfamily 1709.04464}}].

\bibitem{Larkoski_2020}
A.~J. Larkoski, I.~Moult and B.~Nachman, \emph{Jet substructure at the large
  hadron collider: A review of recent advances in theory and machine learning},
  \href{https://doi.org/10.1016/j.physrep.2019.11.001}{\emph{Physics Reports}
  {\bfseries 841} (jan, 2020) 1--63}.

\bibitem{thais2022graph}
S.~Thais, P.~Calafiura, G.~Chachamis, G.~DeZoort, J.~Duarte, S.~Ganguly et~al.,
  \emph{Graph neural networks in particle physics: Implementations,
  innovations, and challenges},  2022.

\bibitem{efn}
P.~T. Komiske, E.~M. Metodiev and J.~Thaler, \emph{Energy flow networks: deep
  sets for particle jets},
  \href{https://doi.org/10.1007/jhep01(2019)121}{\emph{Journal of High Energy
  Physics} {\bfseries 2019} (jan, 2019) }.

\bibitem{ewmp}
P.~Konar, V.~S. Ngairangbam and M.~Spannowsky, \emph{Energy-weighted message
  passing: an infra-red and collinear safe graph neural network algorithm},
  \href{https://doi.org/10.1007/jhep02(2022)060}{\emph{Journal of High Energy
  Physics} {\bfseries 2022} (feb, 2022) }.

\bibitem{particlenet}
H.~Qu and L.~Gouskos, \emph{Jet tagging via particle clouds},
  \href{https://doi.org/10.1103/physrevd.101.056019}{\emph{Physical Review D}
  {\bfseries 101} (mar, 2020) }.

\bibitem{Mikuni:2020wpr}
V.~Mikuni and F.~Canelli, \emph{{ABCNet: An attention-based method for particle
  tagging}}, \href{https://doi.org/10.1140/epjp/s13360-020-00497-3}{\emph{Eur.
  Phys. J. Plus} {\bfseries 135} (2020) 463},
  [\href{https://arxiv.org/abs/2001.05311}{{\ttfamily 2001.05311}}].

\bibitem{Gong:2022lye}
S.~Gong, Q.~Meng, J.~Zhang, H.~Qu, C.~Li, S.~Qian et~al., \emph{{An efficient
  Lorentz equivariant graph neural network for jet tagging}},
  \href{https://doi.org/10.1007/JHEP07(2022)030}{\emph{JHEP} {\bfseries 07}
  (2022) 030}, [\href{https://arxiv.org/abs/2201.08187}{{\ttfamily
  2201.08187}}].

\bibitem{pointnet}
C.~R. Qi, H.~Su, K.~Mo and L.~J. Guibas, \emph{Pointnet: Deep learning on point
  sets for 3d classification and segmentation},  2016.
\newblock 10.48550/ARXIV.1612.00593.

\bibitem{zaheer2018deep}
M.~Zaheer, S.~Kottur, S.~Ravanbakhsh, B.~Poczos, R.~Salakhutdinov and A.~Smola,
  \emph{Deep sets},  2018.

\bibitem{PhysRevLett.41.1581}
G.~C. Fox and S.~Wolfram, \emph{Observables for the analysis of event shapes in
  ${e}^{+}{e}^{-}$ annihilation and other processes},
  \href{https://doi.org/10.1103/PhysRevLett.41.1581}{\emph{Phys. Rev. Lett.}
  {\bfseries 41} (Dec, 1978) 1581--1585}.

\bibitem{mbody}
K.~Datta and A.~Larkoski, \emph{How much information is in a jet?},
  \href{https://doi.org/10.1007/jhep06(2017)073}{\emph{Journal of High Energy
  Physics} {\bfseries 2017} (jun, 2017) }.

\bibitem{Gallicchio_2010}
J.~Gallicchio and M.~D. Schwartz, \emph{Seeing in color: Jet superstructure},
  \href{https://doi.org/10.1103/physrevlett.105.022001}{\emph{Physical Review
  Letters} {\bfseries 105} (jul, 2010) }.

\bibitem{pointnet++}
C.~R. Qi, L.~Yi, H.~Su and L.~J. Guibas, \emph{Pointnet++: Deep hierarchical
  feature learning on point sets in a metric space},  2017.
\newblock 10.48550/ARXIV.1706.02413.

\bibitem{Kingma:2014vow}
D.~P. Kingma and J.~Ba, \emph{{Adam: A Method for Stochastic Optimization}},
  12, 2014, \href{https://arxiv.org/abs/1412.6980}{{\ttfamily 1412.6980}}.

\bibitem{Benato:2021olt}
L.~Benato et~al., \emph{{Shared Data and Algorithms for Deep Learning in
  Fundamental Physics}},
  \href{https://doi.org/10.1007/s41781-022-00082-6}{\emph{Comput. Softw. Big
  Sci.} {\bfseries 6} (2022) 9},
  [\href{https://arxiv.org/abs/2107.00656}{{\ttfamily 2107.00656}}].

\bibitem{Sjostrand:2014zea}
T.~Sj\"ostrand, S.~Ask, J.~R. Christiansen, R.~Corke, N.~Desai, P.~Ilten
  et~al., \emph{{An introduction to PYTHIA 8.2}},
  \href{https://doi.org/10.1016/j.cpc.2015.01.024}{\emph{Comput. Phys. Commun.}
  {\bfseries 191} (2015) 159--177},
  [\href{https://arxiv.org/abs/1410.3012}{{\ttfamily 1410.3012}}].

\bibitem{deFavereau:2013fsa}
{\scshape DELPHES 3} collaboration, J.~de~Favereau, C.~Delaere, P.~Demin,
  A.~Giammanco, V.~Lema\^\i{}tre, A.~Mertens et~al., \emph{{DELPHES 3, A
  modular framework for fast simulation of a generic collider experiment}},
  \href{https://doi.org/10.1007/JHEP02(2014)057}{\emph{JHEP} {\bfseries 02}
  (2014) 057}, [\href{https://arxiv.org/abs/1307.6346}{{\ttfamily 1307.6346}}].

\bibitem{Cacciari:2008gp}
M.~Cacciari, G.~P. Salam and G.~Soyez, \emph{{The anti-$k_t$ jet clustering
  algorithm}}, \href{https://doi.org/10.1088/1126-6708/2008/04/063}{\emph{JHEP}
  {\bfseries 04} (2008) 063},
  [\href{https://arxiv.org/abs/0802.1189}{{\ttfamily 0802.1189}}].

\bibitem{Qu:2019gqs}
H.~Qu and L.~Gouskos, \emph{{ParticleNet: Jet Tagging via Particle Clouds}},
  \href{https://doi.org/10.1103/PhysRevD.101.056019}{\emph{Phys. Rev. D}
  {\bfseries 101} (2020) 056019},
  [\href{https://arxiv.org/abs/1902.08570}{{\ttfamily 1902.08570}}].

\bibitem{Komiske:2018cqr}
P.~T. Komiske, E.~M. Metodiev and J.~Thaler, \emph{{Energy Flow Networks: Deep
  Sets for Particle Jets}},
  \href{https://doi.org/10.1007/JHEP01(2019)121}{\emph{JHEP} {\bfseries 01}
  (2019) 121}, [\href{https://arxiv.org/abs/1810.05165}{{\ttfamily
  1810.05165}}].

\bibitem{Moreno:2019bmu}
E.~A. Moreno, O.~Cerri, J.~M. Duarte, H.~B. Newman, T.~Q. Nguyen, A.~Periwal
  et~al., \emph{{JEDI-net: a jet identification algorithm based on interaction
  networks}}, \href{https://doi.org/10.1140/epjc/s10052-020-7608-4}{\emph{Eur.
  Phys. J. C} {\bfseries 80} (2020) 58},
  [\href{https://arxiv.org/abs/1908.05318}{{\ttfamily 1908.05318}}].

\bibitem{Mikuni:2021pou}
V.~Mikuni and F.~Canelli, \emph{{Point cloud transformers applied to collider
  physics}}, \href{https://doi.org/10.1088/2632-2153/ac07f6}{\emph{Mach. Learn.
  Sci. Tech.} {\bfseries 2} (2021) 035027},
  [\href{https://arxiv.org/abs/2102.05073}{{\ttfamily 2102.05073}}].

\bibitem{Qu:2022mxj}
H.~Qu, C.~Li and S.~Qian, \emph{{Particle Transformer for Jet Tagging}},
  \href{https://arxiv.org/abs/2202.03772}{{\ttfamily 2202.03772}}.

\end{thebibliography}\endgroup
\end{document}